\documentclass [12pt] {article}
\usepackage{graphicx}
\usepackage{epsfig}
\def\lsim{\raise0.3ex\hbox{$<$\kern-0.75em\raise-1.1ex\hbox{$\sim$}}}
\def\gsim{\raise0.3ex\hbox{$>$\kern-0.75em\raise-1.1ex\hbox{$\sim$}}}

\begin {document}
\vskip 0.5cm

\begin{center}
{\Large {\bf QUARK-GLUON STRING MODEL}}\\
\vskip 0.3cm
{\bf {\Large DESCRIPTION OF BARYON PRODUCTION}}\\
\vskip 0.3cm
{\bf {\Large IN {$K^{\pm}$N INTERACTIONS}}} \\

\vskip 1.5 truecm
{ G. H. Arakelyan$^1$, C. Merino$^2$, and Yu. M. Shabelski$^3$}\\
\end{center}

\vskip 1.5 truecm
\begin{center}
{\bf ABSTRACT}
\end{center}

The process of baryon production in $K p$
collisions at high energies is considered in the framework of
the Quark-Gluon String Model.
The  contribution of the string-junction  mechanism to the strange
baryon production is analysed. The results of numerical
calculations  are in reasonable agreement with the data on
inclusive spectra of $p$, $\Lambda$, $\bar{\Lambda}$, and
on the $\bar{\Lambda}/\Lambda$ asymmetry.
The predictions for $\Xi$ and $\Omega$ baryons are presented.

\vskip 1.5 truecm

\noindent $^1$Permanent address: Yerevan Physics Institute,  Armenia\\
E-mail: argev@mail.yerphi.am
\vskip 0.3 truecm

\noindent $^2$Permanent address:  Departamento de F\'\i sica de Part\'\i culas, Facultade de F\'\i sica, and
Instituto Galego de Altas Enerx\'\i as (IGAE),
Universidade de  Santiago de Compostela, Galicia, Spain \\
E-mail: merino@fpaxp1.usc.es

\vskip 0.3 truecm
\noindent $^3$Permanent address: Petersburg Nuclear Physics Institute,
Gatchina, St.Petersburg, Russia\\
E-mail: shabelsk@thd.pnpi.spb.ru

\newpage
\pagestyle{plain}

\noindent{\bf 1. INTRODUCTION}

\vskip 0.3 truecm
The Quark--Gluon String Model (QGSM) is based on the Dual Topological
Unitarization (DTU) and it describes quite reasonably and in a theoretically
consistent way many features of high energy production processes, including 
the inclusive spectra of
different secondary hadrons, their multiplicities, etc., both
in hadron-nucleon and hadron-nucleus collisions \cite{KTM}-\cite{ShK}.
High energy interactions are considered as  taking place via the exchange
of one or several Pomerons, and all elastic and inelastic processes
result from cutting through or between  those exchanged Pomerons \cite{AGK,TMPL44}. The
possibility of different  numbers of Pomerons  to be exchanged  introduces
absorptive corrections to the  cross-sections which are in
agreement with the experimental data on production of hadrons consisting
of light quarks. Inclusive spectra of hadrons are related to the
corresponding fragmentation functions of quarks and diquarks, which
are constructed in terms of the intercepts of well known Regge
trajectories~\cite{Kai}.

In   papers \cite{ACKS}-\cite{AMSh}
the processes connected with the transfer
of baryon charge over long rapidity distances were discussed. In the string models
baryons are considered as configurations consisting of three strings
attached to three valence quarks and connected in one point that it is
called string junction ($SJ$)~\cite{Artru,IOT,RV}. Thus the $SJ$ mechanism has
a nonperturbative origin in QCD.

It is very important to understand the role of the $SJ$ mechanism in
the dynamics of high-energy hadronic interactions, in particular in
processes implying baryon number transfer \cite{Noda}-\cite{Olga}.
Significant results
on this question were obtained in \cite{ACKS,ShBopp,AMSh,Olga}. In these papers
the $SJ$ mechanism was used to analyse the strange baryon production
in $\pi p$ and $pp$ interactions.

The present paper is devoted to the calculation of inclusive spectra of
fast baryon production in the case of kaon beam and to the analysis of the
contribution of $SJ$ mechanism in $Kp$ collisions.

The strange quark in the kaon carries, on the average, a larger fraction of the
momentum than the non-strange quark.
This translates in a harder spectra for the secondary particles produced as result
of the fragmentation of the strange quark.
How much harder these spectra become it will depend on the interaction mechanism,
and therefore the comparison of the theoretical calculations with experiment
offers the possibility of making a more complete test of the model.
The first consideration of the spectra of $\Lambda$-baryon production on
kaon beam \cite{Wright,Barth} was made in~\cite{ShK}, where strange baryon
production was considered in the QGSM scheme
without taking into account the $SJ$ contribution.

In the present paper we analyse the existing data on spectra and
asymmetry of $\Lambda$ and $\bar{\Lambda}$
production on $K$-beams ~\cite{Wright}-\cite{Brick}. We use the same
parametrisations of
diquark fragmentation functions to strange baryons and the same Regge
trajectory intercepts as in \cite{ACKS, ShBopp}.

The obtained description of the experimental data is presented and
new information on the properties of the $SJ$ dynamics is extracted.

In this paper we mainly compare the experimental data with
the result of our calculations for $SJ$ intercept $\alpha_{SJ} = 0.9$.

\begin{center}
\vskip 0.9 truecm
\noindent{\bf 2. INCLUSIVE SPECTRA OF SECONDARY HADRONS \\ IN $Kp$ \-COLLISIONS}
\end{center}
\vskip 0.3 truecm
As it is thoroughly known the exchange of one or several Pomerons is one 
basic feature of high energy hadron-nucleon and hadron-nucleus interactions 
in the frame of QGSM and Dual Parton Model (DPM). Each Pomeron corresponds 
to a cylindrical  diagram, and thus when cutting a Pomeron two showers of 
secondaries are produced. The inclusive spectrum of secondaries is
determined by the convolution of diquark, valence, and sea quark distribution 
functions in the incident particle, $u(x,n)$, and the fragmentation
functions of quarks and diquarks into secondary hadrons, $G(z)$.

The inclusive spectrum (i.e. Feynman-$x$ distribution) of a secondary hadron
$h$ is determined in QGSM by the expression~\cite{KTM}
\begin{equation}
\label{13}
\frac{x_{E}}{\sigma_{inel}}\cdot\frac{d \sigma}{dx_{F}} = \sum_{n = 1}^{\infty}
w_{n}\cdot\varphi_{n}^{h} (x_{F})\;,
\end{equation}
where $x_{E} = E/E_{max}$, and
\begin{equation}
\label{14}
w_{n} = \sigma_{n} \left/ \sum_{n = 1}^{\infty} \sigma_{n} \right.
\end{equation}
is the weight of the diagram with $n$ cutted Pomerons. The $n$ cutted Pomeron 
cross-sections $\sigma_n$ are calculated using the quasi-eikonal
approximation with a supercritical Pomeron~\cite{TMPL44}:
\begin{equation}
\label{r2}
\sigma_n = \frac{\sigma_P}{n\cdot z}\cdot\left (1 - e^{-z}\sum_{k=0}^{\infty}
\frac{z^k}{k!}\right ), \  n \ge 1,
\end{equation}
\begin{equation}
\label{r3}
z = \frac{2C_P\gamma_P}{R^2_P+\alpha'_P\ln(s/s_0)}
\cdot\left(\frac{s}{s_0}\right )^{\Delta},
\end{equation}
\begin{equation}
\label{r2a}
\sigma_P = 8\pi \gamma_P\cdot\left (\frac{s}{s_0}\right )^{\Delta},
\end{equation}
where $\sigma_P$ is the Pomeron contribution to the total cross-section, 
$\Delta  = \alpha_P(0) - 1$ is the excess of the Pomeron intercept over $1$
(supercritical Pomeron), and parameters $\gamma_P$, $R^2_P$, and $C_{p}$ 
take the values for the case of $Kp$ interactions presented in \cite{Volk} 
(see also \cite{ShK}).

The function $\varphi_{n}^{h} (x_{F})$ in Eq.~(\ref{13}) determines the contribution
of the diagram in which $n$ Pomerons
are cut. In the case of $Kp$ collisions this function has the form~\cite{ShK}:
\begin{equation}
\label{Shk1}
\varphi_{n}^{Kp \rightarrow h} (x_{F}) = f_{\overline{q}}^{h} (x_{+},n)\cdot
f_{q}^{h} (x_{-},n) + f_{q}^{h} (x_{+},n)\cdot f_{qq}^{h} (x_{-},n) + 2 (n - 1)
f_{s}^{h} (x_{+},n)\cdot f_{s}^{h} (x_{-},n)\;,
\end{equation}
with
\begin{equation}
\label{Shk2}
x_{\pm} = \frac{1}{2} \left[ \left( \frac{4m_{\perp }^{2}}{s} + x^{2}_{F}
\right)^{ \frac{1}{2}} \pm x_{F} \right]\;.
\end{equation}

The quantities $f_{qq},\; f_{q},\; f_{\overline{q}}$, and $f_{s}$
in Eq.~(\ref{Shk1}) correspond to the contributions of the diquark,
the valence quark and
antiquark, and the sea quarks, while the contributions of the incident
particle and the target proton depend on the variables $x_{+}$ and $x_{-}$,
respectively.

The values of $f_{qq}^{h} (x_{\pm},n),\; f_{q}^{h} (x_{\pm},n),\;
f_{\overline{q}}^{h}(x_{\pm},n)$, and $ f_{s}^{h} (x_{\pm},n)$ can be
obtained through the convolution of the corresponding momentum distribution
of the diquarks, valence quarks, and sea quarks in the colliding hadrons,
$u(x)$, and the function for fragmentation $G^{h}(z)$ of either diquarks
or quarks into secondary hadrons:

\begin{equation}
\label{t3}
f_{q}^{h}(x_{\pm},n) = \int_{x_{\pm}}^{1} u_{q}(x_{1},n)\cdot G_{q}^{h}(x_{\pm}/x_{1})
dx_{1},\ \
\end{equation}

\begin{equation}
\label{t31}
f_{qq}^{h} (x_{-},n) = \frac{2}{3} \int_{x_{-}}^{1} u_{ud} (x_{1},n)\cdot
G_{ud}^{h} (x_{-} / x_{1}) dx_{1} + \frac{1}{3} \int_{x_{-}}^{1} u_{uu}
(x_{1},n)\cdot G_{uu}^{h} (x_{-} / x_{1}) dx_{1} \,
\end{equation}

\begin{eqnarray}
\label{t32}
f_{s}^{h} (x_{\pm},n) & = & \frac{1}{2 + \delta} \left[ \int_{x_{\pm}}^{1}
u_{\overline{u}} (x_{1},n) \frac{G_{\overline{u}}^{h} (x_{\pm} / x_{1}) +
G_{u}^{h} (x_{\pm} / x_{1})}{2} dx_{1} \right. \nonumber  \\
& + & \int_{x_{\pm}}^{1} u_{\overline{d}} (x_{1},n) \frac{G_{\overline{d}}^
{h} (x_{\pm} / x_{1}) + G_{d}^{h} (x_{\pm} / x_{1})}{2} dx_{1} \nonumber  \\
& + & \left. \delta\cdot \int_{x_{\pm}}^{1} u_{\overline{s}} (x_{1},n)
\frac{G_{\overline{s}}^{h} (x_{\pm} / x_{1}) + G_{s}^{h} (x_{\pm} / x_{1})}
{2} dx_{1} \right]. \;
\end{eqnarray}

The parameter $\delta \sim 0.2$-$0.3$ determines here the relative
suppression of strange quarks in the sea. It is important to note that we
use the factor $\delta$ in two cases. The first one is presented in Eq. (10),
where its variation from 0.2 to 0.3 changes the model predictions
very little. The second case is discussed in the Appendix, where $\delta$
determines the relative ratio of strange and non-strange baryon production.
In the latter case the variation of $\delta$ from 0.2 to 0.32
changes the model predictions rather significantly, as it is shown in the
next section.

The detailed description of high energy hadron-nucleon cross-sections on
the base of the Reggeon calculus has been presented in many papers
(see, for instance,~\cite{KTM,KTMS,Sh,KPI,ShK}).

The diquark and quark distribution functions and the fragmentation
functions are determined by the Regge intercepts~\cite{Kai}.

The complete set of distribution and fragmentation functions used in
this paper is presented in the Appendix.

As it was shown in \cite{ACKS} the net baryon charge can be obtained from
the fragmentation of the diquark
giving rise to a leading baryon. \par

\newpage

\vspace{2.cm}
\begin{figure}[htb]
\centering
\includegraphics[width=.55\hsize]{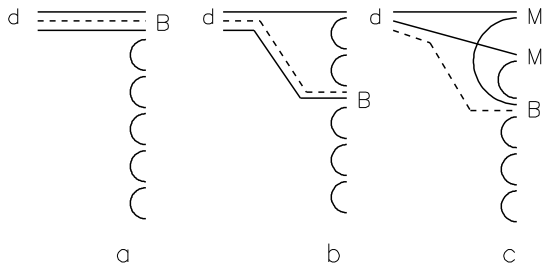}
\caption{QGSM diagrams describing secondary baryon $B$ production
by diquark $d$: (a) initial $SJ$ together with two valence quarks and
one sea quark, (b) initial $SJ$ together with one valence quark and two sea
quarks, and (c) initial $SJ$ together with three sea quarks.}
\end{figure}

To produce a secondary
baryon in the process of diquark fragmentation there exist three possibilities that are shown in Fig.~1.
Thus the secondary baryon can consist of:
the initial $SJ$ together with two valence and one sea quarks (Fig.~1a),
the initial $SJ$ together withone valence
and two sea quarks (Fig.~1b), the initial $SJ$ together with three sea quarks (Fig.~1c). The fraction
of the incident baryon energy carried by the secondary baryon decreases from
case (a) to case (c), whereas the mean rapidity gap between the incident and
secondary baryon increases.

The diagram of Fig. 1b has been used for the description of baryon number
transfer in QGSM \cite{KTM,22r}. Also it describes also the fast meson
production by a diquark~\cite{Kai}.

In this paper we mainly analyse the contribution of
the graph in Fig. 1c to the diquark fragmentation function. This contribution has been
determined for kaon induced reactions in the forward
hemisphere in~\cite{ACKS}. Its magnitude is proportional to one coefficient that it will be
denoted by $\varepsilon$ (the suppression factor of the process of
Fig. 1c compared to those of the processes in Figs. 1a and 1b, see
the Appendix), and that it was firstly analysed in \cite{ACKS,ShBopp,AMSh}
for the case of proton and pion induced reactions.

The $SJ$ mechanism has a nonperturbative
origin and since it is at present not possible to determine $\alpha_{SJ}$ in
QCD from first principles. Thus we  treat  $\alpha_{SJ}$ and $\varepsilon$
as phenomenological parameters which should be determined from experimental
data. In the present calculation, we use the values of parameter
$\alpha_{SJ} = 0.9$ and $\varepsilon=0.024$,
as it was done in \cite{ShBopp,ShBnucl}.
\vskip 0.9 truecm

\noindent{\bf 3. RESULTS AND COMPARISON WITH EXPERIMENT}
\vskip 0.3 truecm

In this paper we mainly consider the existing experimental data on $\Lambda$
and $\bar{\Lambda}$ production on nucleon target~\cite{Wright}-\cite{Brick}.
We present the comparison of QGSM calculations with the experimental data on
spectra and on asymmetry of $\Lambda$ and $\bar{\Lambda}$ hyperons.
We also analyse the role of the strange quark suppression factor $\delta$ (see
the Appendix), and we present results of the calculations for two values of this parameter.
As it was shown in \cite{ShBnucl}, the better agreement of QGSM with data on
strange baryon production on nucleus was obtained with $\delta =0.32$, instead
of the previous value $\delta =0.2$. In principal we cannot exclude the possibility that the
value of $\delta$ could be different for secondary baryons and for mesons
(i.e., for $\Lambda$-baryon and for kaon).

The fragmentation functions into $\bar{\Lambda}$ do not depend on the $SJ$
mechanism, so the $\bar{\Lambda}$ spectra are the same for different values
of $\alpha_{SJ}$, and also they have a very small dependence on the strange quark
suppression factor $\delta$ since they depend on $\delta$ only via Eq.~(10).

\begin{figure}[htb]
\centering
\includegraphics[width=.75\hsize]{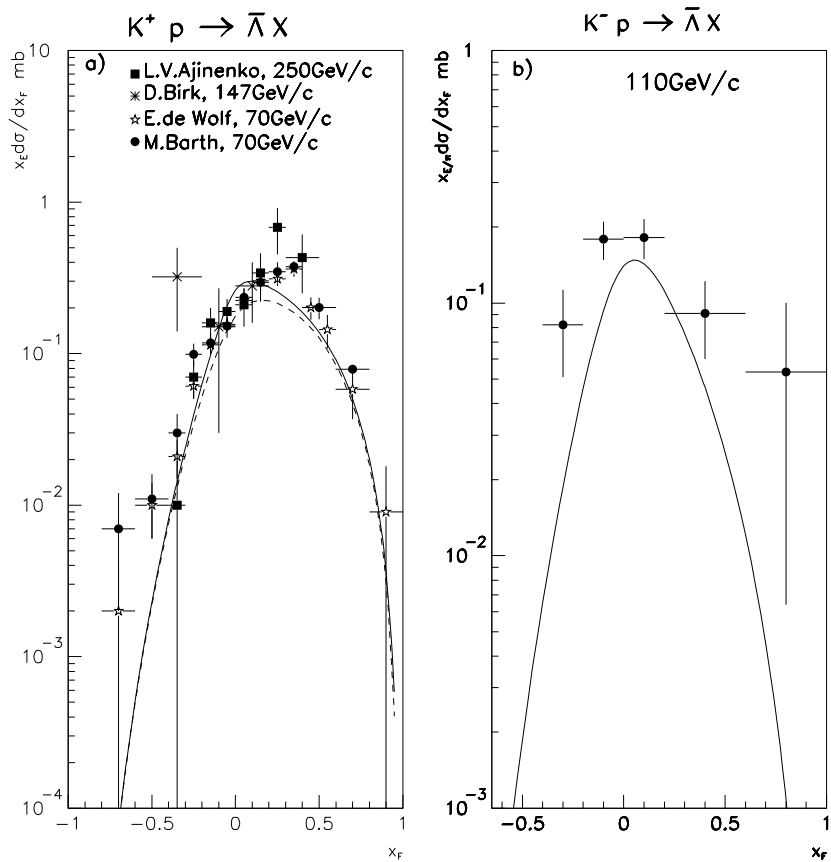}
\caption{The QGSM description of $x_F$ spectra of $\bar{\Lambda}$ in: (a)
$K^+ p$ collisions at 70~GeV/c~\cite{deWolf,Barth}, 147~GeV/c ~\cite{Brick},
and 250~GeV/c~\cite{Ajinenko}. The solid curve corresponds to 250 Gev/c
and the dashed curve to 70 Gev/c. (b) $K^-p$ collisions at
110~GeV/c~\cite{Wright}. The curve shows to the QGSM prediction.}
\end{figure}

The inclusive spectra of $\bar \Lambda$ produced in $K^+ p$ collisions at
$E_{lab}$ = 250 GeV/c \cite{Ajinenko}, 147~GeV/c \cite{Brick}, and 70 GeV/c
\cite{deWolf,Barth} are shown in Fig.~2a. The corresponding spectrum of
$\bar \Lambda$ produced in $K^- p$ collisions at $E_{lab}$ = 110 GeV/c is 
shown in Fig. 2b. The two curves on Fig. 2a correspond to two different values of $E_{lab}$
(the solid curve stands for $E_{lab}$ = 250 GeV/c and the dashed curve for
$E_{lab}$ = 70 GeV/c). As we see there is only very small difference between
these two curves in the central region, so the curve for $E_{lab}$= 147~GeV/c, which lies
in between the other two, is not presented.

\begin{figure}[htb]
\centering
\includegraphics[width=.75\hsize]{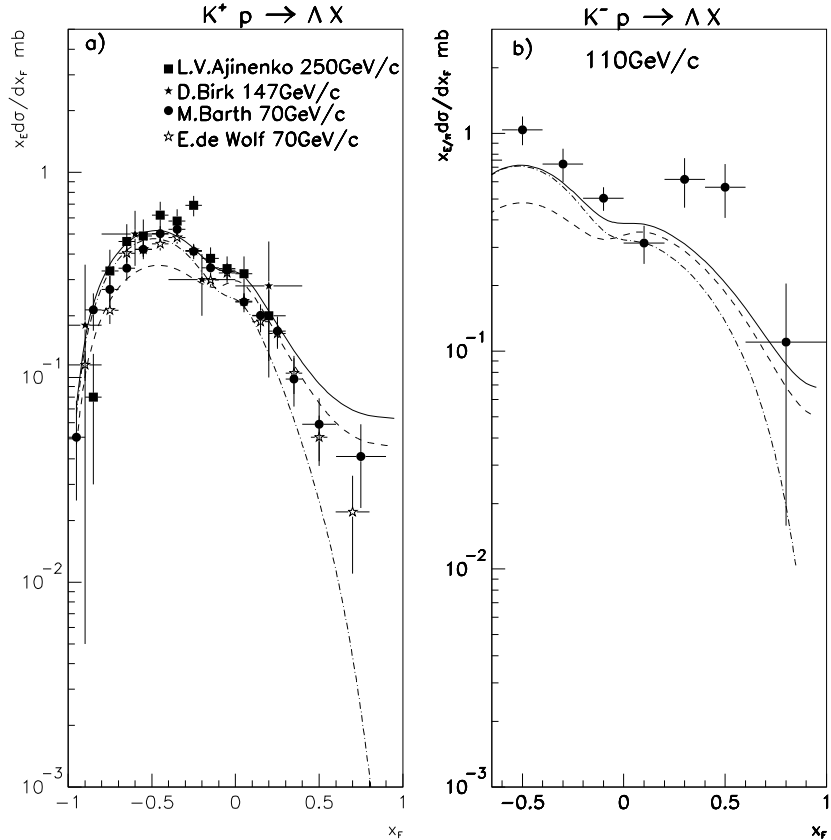}
\caption{The QGSM description of $x_F$ spectra of $\Lambda$ in: (a) $K^+ p$
collisions at 70~GeV/c~\cite{deWolf,Barth}, 147~GeV/c~\cite{Brick}, and
250~GeV/c~\cite{Ajinenko}. (b) $K^-p$ collisions at 110~GeV/c~\cite{Wright}.
The solid curves correspond to $\alpha_{SJ}=0.9$, $\varepsilon = 0.024$, and
$\delta =0.32$, the dashed curves to $\delta =0.2$, and the dashed-dotted
curves to $\varepsilon = 0$.}
\end{figure}
\vskip 0.3 truecm

The inclusive spectra of $\Lambda$ produced in $K^+ p$ collisions
at $E_{lab}$ = 250 GeV/c \cite{Ajinenko}, 147~GeV/c \cite{Brick}, and
70 GeV/c \cite{deWolf,Barth} are shown in Fig.~3a. The corresponding
spectra of $\Lambda$ produced in $K^- p$ collisions
at $E_{lab}=110~GeV/c$ \cite{Wright} are shown on Fig. 3b.

In the following figures the solid curves will correspond to
calculations with $\alpha_{SJ}=0.9$, $\varepsilon = 0.024$, and
$\delta =0.32$, the dashed curves to $\alpha_{SJ}=0.9$, $\varepsilon = 0.024$,
and $\delta =0.2$, and the dashed-dotted curves will stand for calculations with
$\varepsilon = 0$ (i.e., without $SJ$ contribution). In general it seems that the solid curves
are in slightly better agreement with the data, although the
difference between the different curves is not large in the region where
experimental data exist.

In Fig.~4 we show the comparison of the QGSM calculations with the data
on the $\bar\Lambda/\Lambda$ asymmetry produced in $K^+p$ (Fig. 4a) and
$K^-p$ (Fig. 4b) interactions at 250 GeV/c~\cite{Alves}. The
$\bar\Lambda/\Lambda$ asymmetry is defined as
\begin{equation}
\label{t15}
A(\bar{\Lambda}/\Lambda) = \frac{N_{\Lambda} - N_{\bar{\Lambda}}}{N_{\Lambda} + N_{\bar{\Lambda}}}
\end{equation}
for each $x_F$ bin.

The asymmetry data are rather interesting. In the proton fragmentation region the
values of $A(\bar{\Lambda}/\Lambda)$ are close to unity, and that is natural
since a proton fragments into $\Lambda$ with significantly larger
probability than into $\bar{\Lambda}$. In the kaon fragmentation region (at
$x_F$ values where experimental data exist) $A(\bar{\Lambda}/\Lambda)$ becomes
negative and decreases very fast in the case of $K^+$ beam. In the case of
$K^-$ beam it increases very fast with $x_F$ and the difference between the
calculations with different parameters is rather small. Both these
behaviors are
also natural, because the $K^+$ contains a $\bar{s}$ valence quark which preferably
fragments into $\bar{\Lambda}$, while the valence $s$ quark from the $K^-$
fragments rather often into $\Lambda$. However, in both cases
the $A(\bar{\Lambda}/\Lambda)$
expermental $x_F$-dependences are much steeper than the
theoretical predictions. This is a probable indication that the fragmentation
functions $s \to \Lambda$ and  $\bar{s} \to \bar{\Lambda}$ should be
further enhanced.

In the $K^+$ fragmentation region (Fig. 4a) the predicted values of
$A(\bar{\Lambda}/\Lambda)$ at $x_F > 0.4$ show a change of behavior and they start
increasing. In this region the contribution of the direct fragmentation of
$\bar{s} \to \bar{\Lambda}$, which makes $A(\bar{\Lambda}/\Lambda)$ to decrease,
becomes smaller than the effect of $SJ$ diffusion which increases the
multiplicity of $\Lambda$. The measurement of the asymmetry
$A(\bar{\Lambda}/\Lambda)$ in the region $x_F \geq 0.4$ in $K^+p$
collisions could make the situation more clear.
\vskip 0.3 truecm
\begin{figure}[htb]
\centering
\label{ass}
\includegraphics[width=.75\hsize]{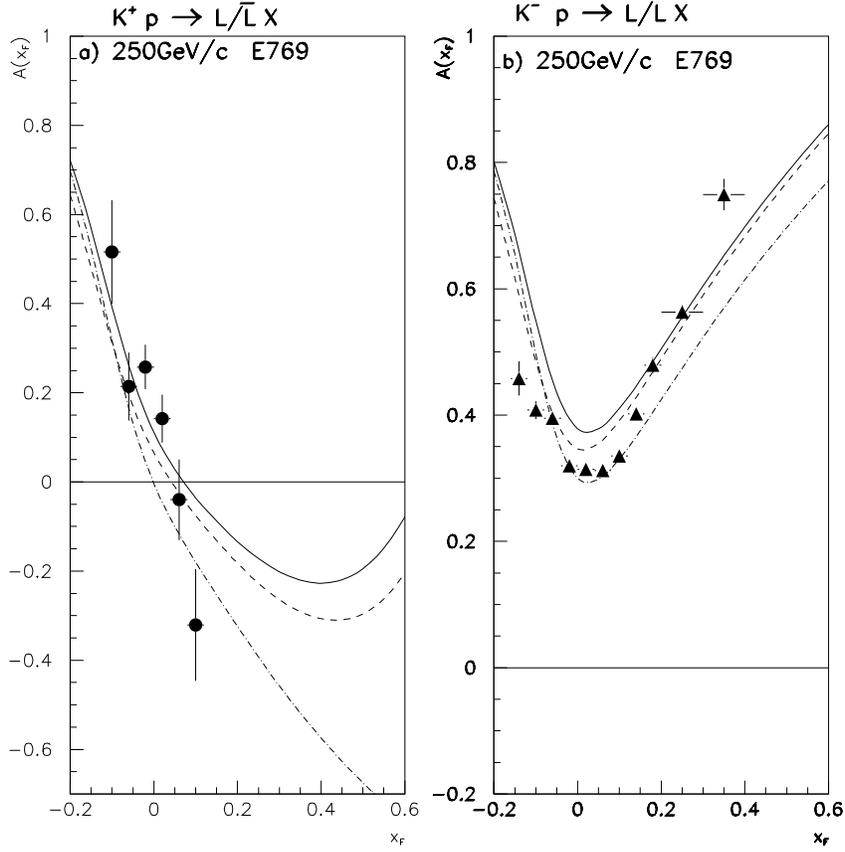}
\caption{The $\bar{\Lambda}/\Lambda$ asymmetry in (a) $K^+p$ and (b) $K^-p$
collisions. Experimental data at 250~GeV/c~\cite{Alves} and the corresponding
QGSM description. The solid curve corresponds to $\alpha_{SJ}=0.9$,
$\varepsilon = 0.024$, and $\delta =0.32$, the dashed curve to $\delta =0.2$,
and the dashed-dotted curve to $\varepsilon =0.$}
\end{figure}
\vskip 0.3 truecm

The inclusive spectra of secondary protons  produced in $K^+p$ collisions at
energies  $E_{lab}$ = 250 GeV/c \cite{Ajinenko}, together with the theoretical
curves, are presented in Fig.~5.
Unfortunately the data on proton production on $K$ beams were only measured
in the backward hemisphere where they are practically the same that in the
case of the pp interaction, with the evident difference in normalization.

\begin{figure}[htb]
\centering
\includegraphics[width=.75\hsize]{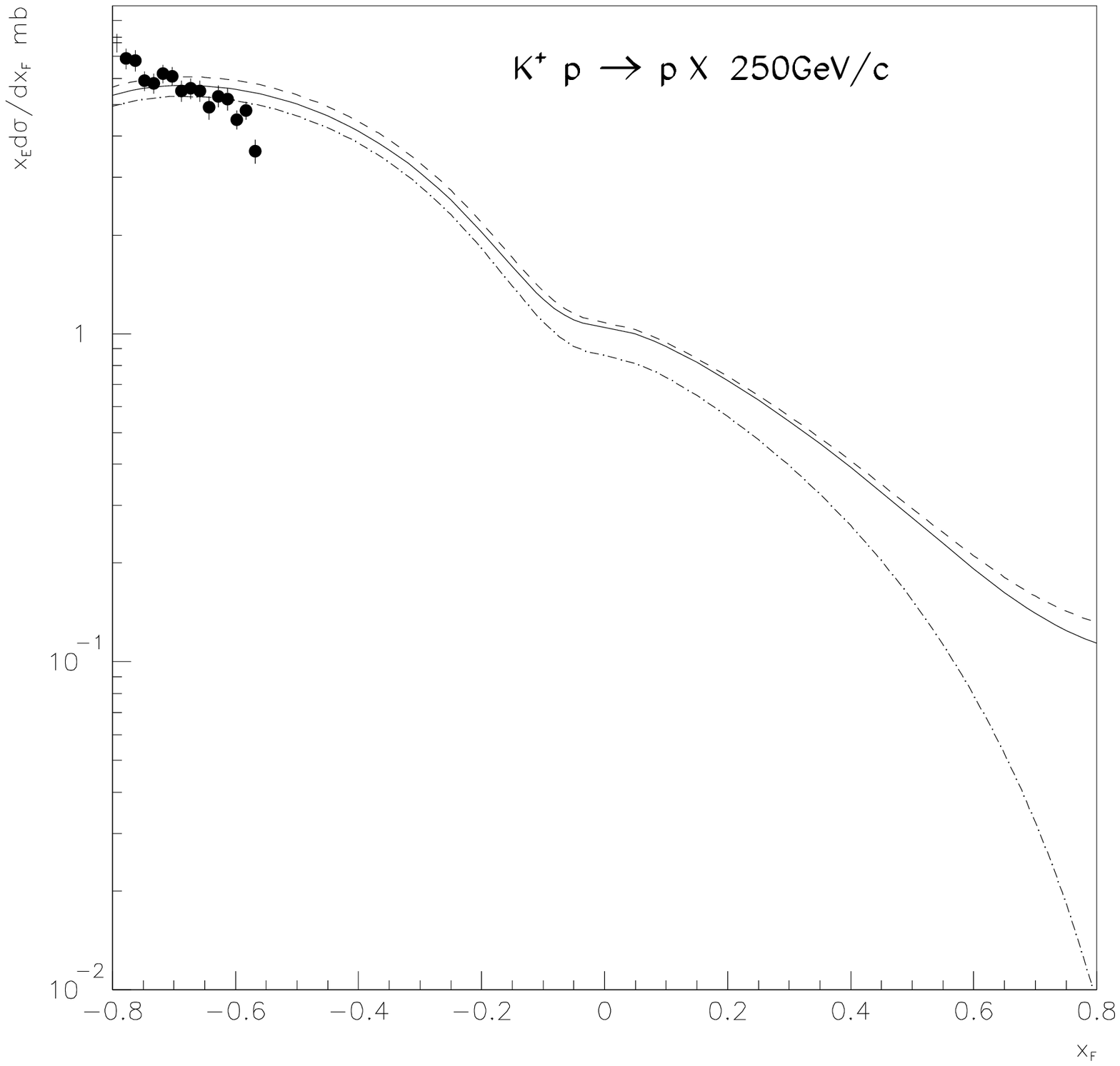}
\caption{The $x_F$-dependence of the spectra of protons in $K^+ p$
collisions. Experimental data at 250~GeV/c~\cite{Ajinenko} and the 
corresponding QGSM description. The full line corresponds to
$\alpha_{SJ}=0.9$, $\varepsilon = 0.024$, and $\delta =0.32$, the
dashed line to $\delta =0.2$, and the dashed-dotted line to $\varepsilon =0.$}
\end{figure}
\vskip 0.3 truecm

In Figs. 6 and 7 we present the predictions for cross sections (Fig. 6) and
asymmetries (Fig. 7) for $\Xi^-$ and $\Omega$ production in $K^+ p$ and
$K^- p$ collisions at 250~GeV/c. In the central region of $K^+p$ collisions
the yields of $\Xi^-$ and $\bar \Xi^+$, as well as of $\Omega^-$ and
$\bar \Omega^+$ are predicted to be practically the same. The smaller
fragmentation function of valence $\bar{s}$ quark into strange baryon is
compensated by the larger fragmentation function of the target diquark.

The large difference between $\Omega$ and $\bar{\Omega}$ production cross-sections
in $K^- p$ collisions (Fig. 6b) can be explained by the presence of valence $s$
quark in $K^-$ meson directly fragmenting to $\Omega$ baryon, and by absence
of valence quarks for the case of $\bar \Omega$ production. For $\Xi^-$ and $\bar \Xi^+$
production one also has valence $s$ quark going into the leading $\Xi^-$ baryon.
So, as it can be seen in Fig. 6b, the cross-sections for $\Xi^-$ and $\Omega$ are
larger than the corresponding cross-sections for antibaryons.

In $K^+ p$ collisions one has the valence $\bar s$ quark to produce leading
$\bar \Xi^+$ and $\bar \Omega$. Thus up to $x \sim 0.5$, these cross sections
are larger than those for producing $\Xi^-$ and $\Omega$. At larger $x_F$ ($x_F\rightarrow 1$) the
antibaryons cross-sections fall more rapidly.

\begin{figure}[htb]
\centering
\label{pcksiom}
\includegraphics[width=.75\hsize]{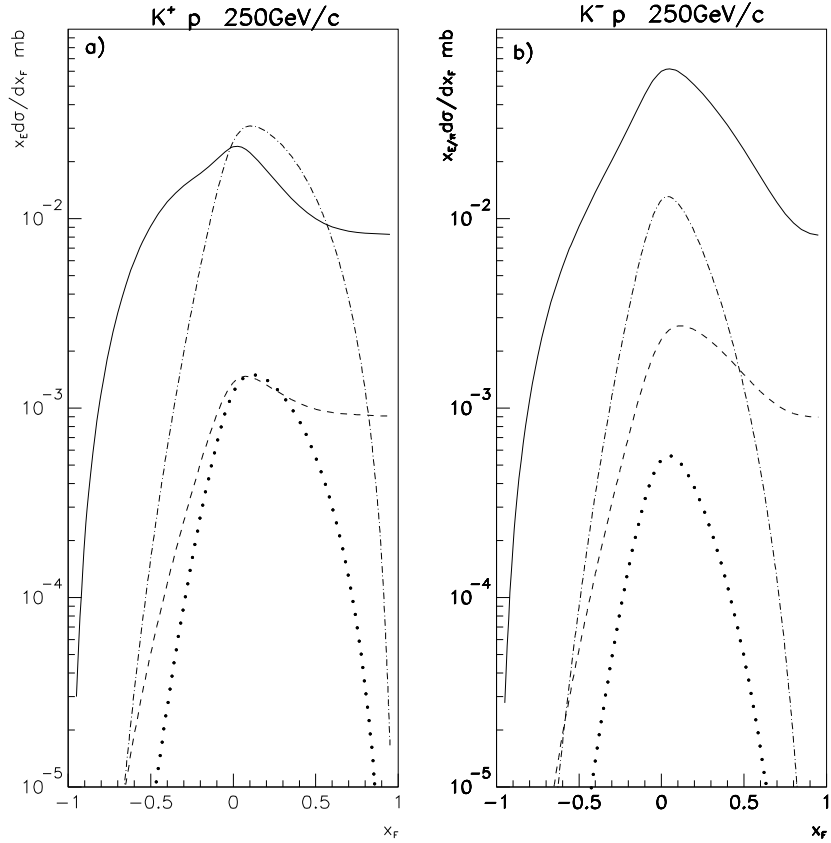}
\caption{QGSM prediction for the $x_F$-dependence of the spectra of heavy
strange baryons in (a) $K^+ p$ and (b) $K^- p$ collisions at 250~GeV/c.
Solid curves are for $\Xi^-$, dash-dotted curves for $\bar{\Xi}^+$,
dashed curves for $\Omega^-$, and dotted curves for $\bar{\Omega}^+$.}
\end{figure}

The predictions for asymmetry in $\Xi$ and $\Omega$ baryon production in
$K^+p$ and $K^-p$ interactions are presented in Fig. 7. Here the general situation
is similar to that of the case of asymmetry in $\Lambda$ production shown in
Fig. 4. In the case of $K^-$ beam the asymmetry for both $\Xi$ and $\Omega$
productions increases in the whole region of positive $x_F$.

\vskip 0.3 truecm

\begin{figure}[htb]
\centering
\label{asksiom}
\includegraphics[width=.75\hsize]{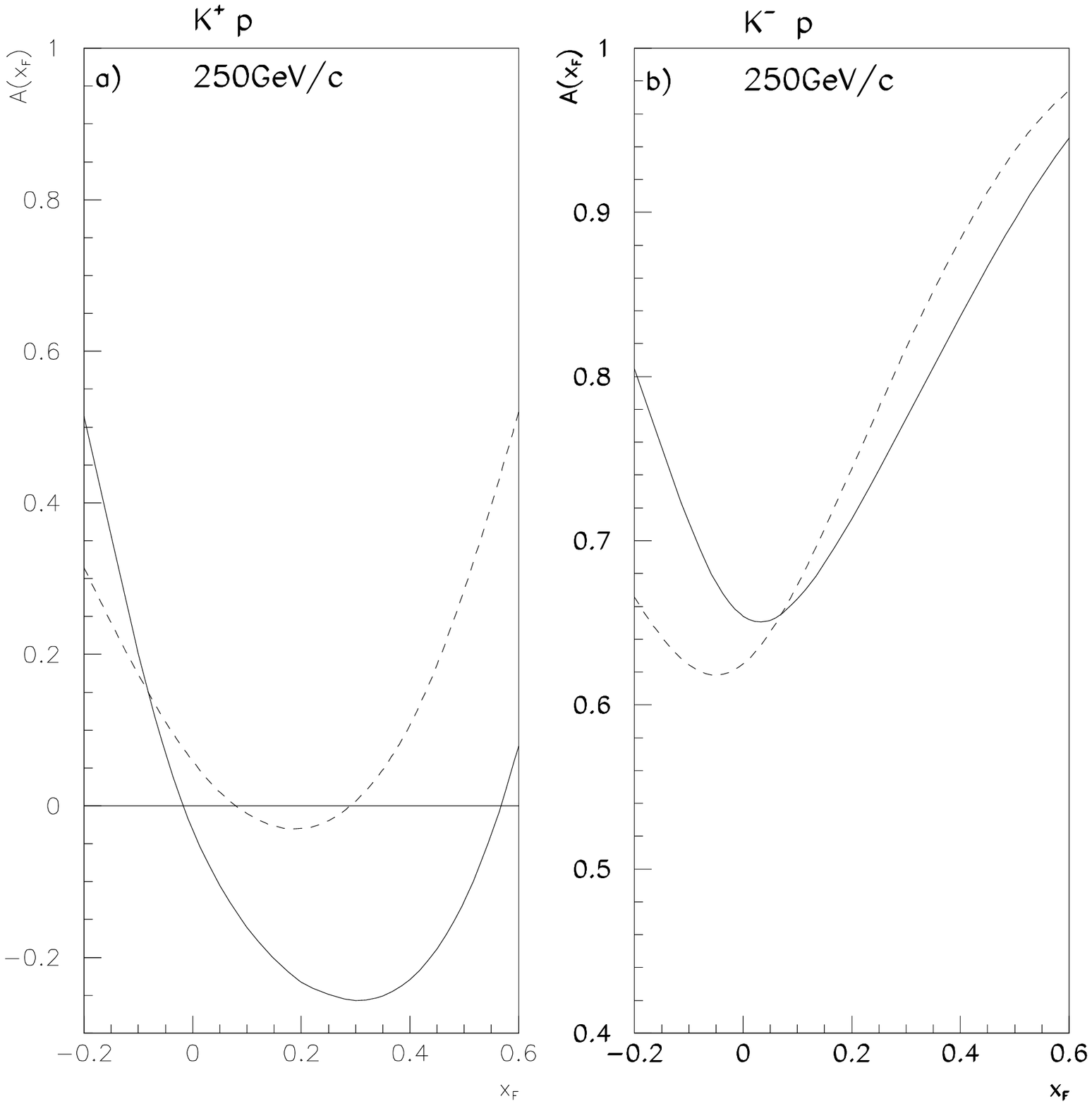}
\caption{QGSM prediction for the $x_F$-dependence of the asymmetry of
heavy strange baryons in (a) $K^+p$ and (b) $K^-p$ collisions at 250~GeV/c.
Solid curves are for $\Xi^-$,dashed curves for $\Omega^-$}
\end{figure}

\vskip 0.9 truecm
\noindent{\bf 5. CONCLUSIONS}
\vskip 0.3 truecm

It is shown that the experimental data can not discriminate among the
different sets of parameters in the model. The results of the calculations with the value
$\alpha_{SJ}=0.5$ do not differ from the results with $\alpha_{SJ}=0.9$ in any
significant way. Even the value $\varepsilon = 0$ seems compatible with the
data. For $K^+$ beam all curves show a more or less reasonable agreement with
experimental data on cross-sections and asymmetry (Figs. 2a and 3a). The
difference between the results for the two values of the strangeness
suppression factor $\delta$ is small in the backward hemisphere but it
appears evident when $x_F$ is positive both for cross-sections and
asymmetries. Unfortunately, the data on asymmetry are concentrated in the
central region  $-0.2<x_F<~0.1$ where the difference among curves is not so
large.

The situation for $K^-$ beam seems worse than in the $K^+$ case since all
curves are, as a rule, below the experimental data. Both the number and
the quality of experimental data on $K^-$ beam are not high enough.

One has also to note that the QGSM does not reproduce the experimentally
observed maximum in the $\Lambda$ spectrum in the range $x_{F} = 0.2$ - $0.4$
which is seen in Fig. 3b (there are no analogous maxima in the reactions
$\pi^{-}p\rightarrow\overline{p}X$ and
$\pi^{+}p \rightarrow pX$~\cite{Brenner}).

QGSM predicts a weak energy dependence of the $\Lambda$ and $\bar{\Lambda}$
production cross-section in $Kp$ collisions at the considered energies.

Finally, the experimental data on high-energy $\Lambda$ production are not
in contradiction with the possibility of baryon charge transfer over large
rapidity distances. The $\bar{\Lambda}/\Lambda$ asymmetry is provided by
$SJ$ diffusion through baryon charge transfer.

The presence of baryon asymmetry in the projectile hemisphere for $Kp$
collisions provides good evidence for such a mechanism.

To get a good understanding of the dynamics of the baryon charge transfer
over large rapidity distances new experimental data in meson and baryon
collisions with nucleons and nuclear targets are needed.

\newpage
\vskip 0.5 truecm
\noindent {\bf Acknowledgments}
\vskip 0.2 truecm

The authors are thankful to A.B.~Kaidalov and C.~Pajares for useful
discussions.

This work was partially financed by CICYT of Spain through contract
FPA2002-01161, and by Xunta de Galicia through contract PGIDIT03PXIC20612PN.

G.~H.~Arakelyan and C.~Merino were also supported by NATO grant CLG.980335 and
Yu.~M.~Shabelski by grants NATO PDD (CP) PST.CLG 980287, RCGSS-1124.2003.2.
G.H.A. thanks the Xunta de Galicia for financial support.

G.H.A and Yu.M.Sh.
want to thank the members of the Department of Particles Physics and of the
Instituto Galego de Altas Enerx\'\i as~(IGAE), University of Santiago de Compostela,
Galicia, Spain, for their kind hospitality
during the final stage of this work.

\newpage
\vskip 0.9 truecm \noindent{\bf APPENDIX: QUARK AND DIQUARK DISTRIBUTIONS AND
THEIR FRAGMENTATION FUNCTIONS}
\vskip 0.3 truecm

The distribution functions of the valence quarks ($u$ quark for $K^+$ and
$s$ quark for $K^-$) in the $K$ meson were chosen in the form \cite{ShK}:
\begin{eqnarray}
\label{u1}
u_{s}^{K^-}(x,n) =u_{\overline {s}}^{K^+}(x,n)= C_{q} x^{-\alpha_{\varphi}} (1 - x)^{-\alpha_{R} + n - 1} \;,\\
u_{\overline{u}}^{K^-}(x,n) = u_{u}^{K^+}(x,n) = C_{\overline{q}} x^{-\alpha_{R}} (1 - x)^
{-\alpha_{\varphi} + n - 1} \;, \nonumber
\end{eqnarray}
for valence quarks, and
\begin{eqnarray}
u(x,n) =  Cx^{-\alpha_R}(1-x)^{n-\alpha_R -1}[1-\delta \sqrt{1-x}]
   \; , \; n>1 \; ,\\
u_{s}(x,n) = C_{s}x^{-\alpha_R}(1-x)^{n-1} \; , \; n>1\ \; ,
\end{eqnarray}
for sea quarks.

The strange valence quark in the $K^-$ meson (and the strange valence antiquark
in the $K^+$ meson) carries on the average twice as
much momentum as the nonstrange antiquark.

The $\delta$ is the relative probability to find a strange quark in
the sea, while the normalization factors $C_{i}$ are determined by the condition
\begin{equation}
\int_{0}^{1} u_{i}(x,n)dx = 1\ \; ,
\end{equation}
and the sum rule
\begin{equation}
\int_{0}^{1}\sum_i u_{i}(x,n)xdx = 1\ \
\end{equation}
is fulfilled.

The quark and diquark distribution functions in the proton have been taken
as in~\cite{ACKS}.

The fragmentation functions of quarks and diquarks have been slightly changed
with respect to Refs.~\cite{Sh,KPI} in order to obtain a better agreement
with the existing experimental data (the fragmentation functions
in~\cite{Sh,KPI} correspond to $\varepsilon$ = 0). We have used the following
functional form for the quark fragmentation functions:
\begin{equation}
\label{g1}
G_{u}^{p} = G_{d}^{p} =
a_{\bar{N}}(1-z)^{\lambda + \alpha_R - 2 \alpha_B}\cdot (1+a_{1}z^{2}) \;\;,
G_{u}^{\bar{p}} = G_{d}^{\bar{p}} = (1-z)\cdot G_{u}^{p},
\end{equation}
\begin{equation}
\label{t4}
G_{u}^{\Lambda} = G_d^{\Lambda} = {a_{\bar{\Lambda}}}
(1-z)^{\lambda + \alpha_R - 2 \alpha_B+\Delta \alpha }\cdot (1+a_{1}z^{2}) \;\;,\\
 G_u^{\bar{\Lambda}} = G_{\bar{d}}^{\Lambda}= (1-z)\cdot G_d^{\Lambda},\nonumber
\end{equation}
\begin{equation}
G_d^{\Xi^-} = \frac{a_{\bar{\Xi}}}{a_{\bar{\Lambda}}}
(1-z)^{\Delta \alpha} G_u^{\Lambda} \;\; ,
G_u^{\Xi^-} = G_u^{\bar{\Xi}} = (1-z) G_d^{\Xi^-}\; ,
\end{equation}
\begin{equation}
G_{u}^{\Omega} = G_d^{\Omega} = G_u^{\bar{\Omega}} = G_d^{\bar{\Omega}}
= \frac{a_{\bar{\Omega}}}{a_{\bar{\Xi}}}
(1-z)^{\Delta \alpha} G_u^{\Xi} \;\; ,
\end{equation}
\begin{equation}
\label{tp}
G_{s}^{p}(z) =  a_{N}(1 - z)^{\lambda +\alpha_{R} -2\cdot \alpha_{B}+1.5}\cdot (1 + a_{1}z)\;,
\end{equation}
\begin{equation}
\label{tap}
G_{s}^{\bar{p}}(z) =  a_{\bar{N}}(1 - z)^{\lambda +\alpha_{R} -2\cdot \alpha_{B}+0.5}\cdot (1+a_{1}z)\;,
\end{equation}
\begin{equation}
\label{ts}
G_{s}^{\Lambda}(z) = a_{\bar{N}}(1 - z)^{\lambda +
\alpha_{R}-2\cdot \alpha_{B}}\cdot (1 + a_{1}z) \;,
\end{equation}
\begin{equation}
\label{tas}
G_{s}^{\bar{\Lambda}}(z) = a_{\bar{\Lambda}}(1 - z)^{\lambda+ \alpha_{R}
-2\alpha_{B} + 2(1 - \alpha_{R}) + 2\Delta \alpha}\;,
\end{equation}
\begin{equation}
\label{tksi}
G_{s}^{\Xi^-}(z) = a_{\overline{\Lambda}}(1 - z)^{\lambda+ \alpha_{R}
-2\alpha_{B} + \Delta \alpha}\cdot (1 + a_{1}z)\;,
\end{equation}
\begin{equation}
\label{taksi}
G_{s}^{\bar \Xi^-}(z) = a_{\overline{\Xi}}(1 - z)^{\lambda+ \alpha_{R}
-2\alpha_{B} + \Delta \alpha +2(\alpha_{R} - \alpha_{\phi})}\cdot (1 + a_{1}z)\;,
\end{equation}
\begin{equation}
\label{taomega}
G_{s}^{\bar\Omega}(z) = a_{\overline{\Omega}}(1 - z)^{4\lambda+ \alpha_{R}
-2\alpha_{B} + 1.}\cdot (1 + a_{1}z)\;, and
\end{equation}
\begin{equation}
\label{taso}
G_{s}^{\Omega}(z) = a_{\overline{\Xi}}(1 - z)^{3\lambda+ \alpha_{R}
-2\alpha_{B}}\cdot (1 + a_{1}z)\;,
\end{equation}
with
\begin{equation}
\label{t5}
\alpha_{R} = 0.5, \alpha_{\Phi} = 0,  \alpha_B = -0.5,
\Delta \alpha = \alpha_{R} - \alpha_{\phi},
\lambda=2\alpha^{\prime}\cdot <p_{t}^2>=0.5 \;.\nonumber
\end{equation}

The fragmentation function of $s$ quark into $\Lambda$ in Eq.~\ref{ts}
differs from the one in~\cite{ShK}. We propose that the fragmentation
functions of strange $s$ quark and light $u$ quark into $\Lambda$ describe the
same mechanism: in both cases the fragmented quark captures the $ud$ diquark
with the corresponding coefficient being the same, $a_{\bar{N}}$. For the
fragmentation of $s$ quark into $\bar{\Lambda}$ we should change the strange quark 
by the strange antiquark so the exchange of one $s \bar s$ system occurs.

The parametrisation of diquark fragmentation functions taking into account
the $SJ$ contribution was analysed in previous papers \cite{ACKS,ShBopp,AMSh},
in which two values of the $SJ$ intercept $\alpha_{SJ} = 0.5$ and
$\alpha_{SJ} = 0.9$ \cite{AMSh} were considered. In the present paper we
have used the parametrisations of fragmentation functions of diquarks into
strange baryons for $\alpha_{SJ} = 0.9$ and the corresponding set of parameters,
as it has been done in~\cite{ShBopp,ShBnucl}. These rather large value of
$\alpha_{SJ}$ is confirmed by high energy data from HERA and RHIC
\cite{ShBopp}.

Diquark fragmentation functions have more complicated forms. They contain
two contributions. The first one corresponds to the central
production of a $B\bar{B}$ pair and it can be described by the
formulas:

\begin{equation}
\label{t6}
G_{uu}^p = G_{ud}^p = G_{uu}^{\bar{p}} = G_{ud}^{\bar{p}}
= a_{\bar{N}}(1-z)^{\lambda-\alpha_R + 4(1-\alpha_B)} \ \ ,
\end{equation}
\begin{equation}
\label{t7}
G_{uu}^{\Lambda} = G_{ud}^{\Lambda} = G_{uu}^{\bar{\Lambda}}
= G_{ud}^{\bar{\Lambda}} = {a_{\bar{\Lambda}}}(1-z)^{\Delta \alpha} G_{uu}^p \;\; ,
\end{equation}
\begin{equation}
G_{uu}^{\Xi^-} = G_{ud}^{\Xi^-} = G_{uu}^{\bar{\Xi}} = G_{ud}^{\bar{\Xi}}
= a_{\bar{\Xi}}(1-z)^{\Delta \alpha}
G_{uu}^{\Lambda} \;\; , and
\end{equation}
\begin{equation}
G_{uu}^{\Omega} = G_{ud}^{\Omega} = G_{uu}^{\bar{\Omega}}
= G_{ud}^{\bar{\Omega}}
= a_{\bar{\Omega}} (1-z)^{\Delta \alpha}
G_{uu}^{\Xi},
\end{equation}
with the same $\Delta \alpha$ as in Eq.~(\ref{t5}).

The second contribution is connected with the direct fragmentation of the
initial baryon into the secondary one with conservation of $SJ$.
As it was discussed above, there exist three
different types of such contributions (Figs.~1a-1c). These contributions
are determined by expressions similar to Eqs.~(\ref{t3}- \ref{t32})
with the corresponding fragmentation functions given by
\begin{equation}
\label{t13}
G_{uu}^p = G_{ud}^p = a_N z^{\beta}\Big[v_0\varepsilon (1-z)^2 +
v_q z^{2 - \beta} (1-z) + v_{qq}z^{2.5 - \beta}\Big] \; ,
\end{equation}
\begin{equation}
\label{t14}
 G_{ud}^{\Lambda} = a_N z^{\beta}
\Big[v_0\varepsilon (1-z)^2 + v_q z^{2 - \beta} (1-z) +
v_{qq}z^{2.5 - \beta}\Big](1-z)^{\Delta \alpha}\; , \;
G_{uu}^{\Lambda} =(1-z)G_{ud}^{\Lambda} \;,
\end{equation}
\begin{equation}
G_{d,SJ}^{\Xi^-} = a_N z^{\beta} [v_0\varepsilon (1-z)^2 +
v_q z^{3/2} (1-z)] (1-z)^{2\Delta \alpha}, ~G_{u,SJ}^{\Xi^-}=
(1-z)G_{d,SJ}^{\Xi^-}\; ,
\end{equation}
\begin{equation}
G_{SJ}^{\Omega} = a_N v_0\varepsilon z^{\beta}
(1-z)^{2+3\Delta \alpha} \; ,
\end{equation}
with $\beta = 1 - \alpha_{SJ}$. As for the factor $z^{\beta}\cdot z^{2-\beta}$
in the second term, it~is~$2(\alpha_R - \alpha_B)$~\cite{KTM}. For the third
term we have added an extra factor $z^{1/2}$.

The probabilities of transition into the secondary baryon of $SJ$ without
valence quarks, $I_3$, $SJ$ plus one valence quark, $I_2$, and $SJ$ plus a
valence diquark, $I_1$, are taken from the simplest quark combinatorials~\cite{CS}.
Assuming that the strange quark suppression is the same in all
these cases, we obtain for the relative yields of
different baryons from $SJ$ fragmentation without valence quarks:
\begin{equation}
\label{t8}
I_3 = 4L^3 : 4L^3 : 12L^2S : 3LS^2 : 3LS^2 : S^3\; ,
\end{equation}
for secondary $p$, $n$, $\Lambda + \Sigma$, $\Xi^0$, $\Xi^-$, and
$\Omega$, respectively.

For $I_2$ we obtain
\begin{equation}
\label{t9}
I_{2u} = 3L^2 : L^2 : 4LS : S^2 : 0\; ,
\end{equation}
and
\begin{equation}
\label{t10}
I_{2d} = L^2 : 3L^2 : 4LS : 0 : S^2\; 
\end{equation}
for secondary $p$, $n$, $\Lambda + \Sigma$, $\Xi^0$ and  $\Xi^-$.

For $I_1$ we have
\begin{equation}
\label{t11}
I_{1uu} = 2L : 0 : S
\end{equation}
and
\begin{equation}
\label{t12}
I_{1ud} = L : L : S
\end{equation}
for secondary $p$, $n$, and $\Lambda + \Sigma$. The ratio $\delta = S/L$
determines the strange suppression factor, and $2L + S$ = 1.
In the numerical calculations we have compared the results for two different 
values $\delta=0.2$ and $\delta=0.32$.

In agreement with the empirical rule we assume that
$\Sigma^+ + \Sigma^- = 0.6\Lambda$ \cite{appelsh} in
Eqs.~(\ref{t8})-(\ref{t10}).
As it is customary $\Sigma^0$ are included into $\Lambda$. Note that this empirical
rule, used in many experimental papers, is not consistent with the simplest
quark statistics rules \cite{AKNS}.

The values of $v_0$, $v_q$, and $v_{qq}$ are directly determined by the
corresponding coefficients in Eqs.~(\ref{t8})-(\ref{t12}), together with the
probabilities to fragment a $qqs$ system into $\Sigma^+ + \Sigma^-$ and
into $\Lambda$, given above. Thus we have: \\
for incident $uu$ diquark and secondary proton
\begin{equation}
\label{vuup}
v_0 = 4L^3 \;\;, v_q = 3L^2 \;\;,v_{qq} = 2L \;,
\end{equation}
for incident $ud$ diquark and secondary proton
\begin{equation}
\label{vudp}
v_0 = 4L^3 \;\;, v_q = 2L^2 \;\;,v_{qq} = L \;,
\end{equation}
for incident $uu$ diquark and secondary $\Lambda$
\begin{equation}
\label{vuul}
v_0 = \frac{12}{1.6}S L^2 \;\;, v_q = \frac4{1.6}S L \;\;,
v_{qq} = \frac14 S \;,
\end{equation}
for incident $ud$ diquark and secondary $\Lambda$
\begin{equation}
\label{vudl}
v_0 = \frac{12}{1.6}S L^2 \;\;, v_q = \frac4{1.6}S L \;\;,
v_{qq} = S \;,
\end{equation}
for incident $u$ quark and secondary $\Xi^-$
\begin{equation}
v_0 = 3 S^2 L \;\;, v_q = 0 \;,
\end{equation}
for incident $d$ quark and secondary $\Xi^-$
\begin{equation}
v_0 = 3 S^2 L \;\;, v_q = S^2 \;,
\end{equation}
and for incident $SJ$ and secondary $\Omega^-$
\begin{equation}
v_0 = S^3 \;.
\end{equation}

The model parameters for quark and diquark distribution functions and their
corresponding fragmentation functions have been mainly taken from the description of the data
in ~\cite{ACKS,ShBopp}.

\end{document}